\documentclass[fleqn,usenatbib]{mnras}

\usepackage{newtxtext,newtxmath}

\usepackage[T1]{fontenc}

\DeclareRobustCommand{\VAN}[3]{#2}
\let\VANthebibliography\thebibliography
\def\thebibliography{\DeclareRobustCommand{\VAN}[3]{##3}\VANthebibliography}

\usepackage{graphicx}	
\usepackage{amsmath}	

\usepackage[dvipsnames]{xcolor} 

\title[Impact of stellar evolution on snowline locations]{The impact of pre-main sequence stellar evolution on midplane snowline locations and C/O in planet forming discs}

\author[J. M. Miley et al.]{
James M. Miley$^{1,2,3}$\thanks{E-mail: james.miley@alma.cl},
Olja Pani\'c$^{1}$,
Richard A. Booth$^{4,5}$,
John D. Ilee$^{1}$,
Shigeru Ida$^{6}$ 
\newauthor and Masanobu Kunitomo $^{7}$
\\
$^{1}$ School of Physics \& Astronomy, University of Leeds, Woodhouse Lane, Leeds,  LS2 9JT, UK\\
$^{2}$ Joint ALMA Observatory, Alonso de Cordova 3107, Vitacura, Santiago, Chile\\
$^{3}$  National Astronomical Observatory of Japan, Alonso de Cordova 3788, 61B Vitacura, Santiago, Chile\\
$^{4}$ Institute of Astronomy, University of Cambridge, Madingley Road, Cambridge, CB3 0HA, UK\\
$^{5}$ Imperial College London, Blackett Laboratory, Prince Consort Road, London, SW7 2AZ, UK\\
$^{6}$ Earth-Life Science Institute, Tokyo Institute of Technology, 2-12-1 Ookayama, Meguro-ku, Tokyo 152-8551, Japan \\
$^{7 }$ Department of Physics, School of Medicine, Kurume University, 67 Asahimachi, Kurume, Fukuoka 830-0011, Japan 
}

\date{Accepted 2020 November 2. Received 2020 October 30; in original form 2020 July 9}

\pubyear{2020}

\begin{document}
\label{firstpage}
\pagerange{\pageref{firstpage}--\pageref{lastpage}}
\maketitle

\begin{abstract}

We investigate the impact of pre-main sequence stellar luminosity evolution on the thermal and chemical properties of disc midplanes. We create template disc models exemplifying initial conditions for giant planet formation for a variety of stellar masses and ages. These models include the 2D physical structure of gas as well as 1D chemical structure in the disc midplane.
The disc temperature profiles are calculated using fully physically consistent radiative transfer models for stars between 0.5 and 3~M$_\odot$ and ages up to 10~Myr. The resulting temperature profiles are used to determine how the chemical conditions in the mid-plane change over time. We therefore obtain gas and ice-phase abundances of the main carbon and oxygen carrier species.
While the temperature profiles produced are not markedly different for the stars of different masses at early stages ($\leq$1~Myr), they start to diverge significantly beyond 2~Myr. Discs around stars with mass $\geq$1.5~M$_\odot$ become warmer over time as the stellar luminosity increases, whereas low-mass stars decrease in luminosity leading to cooler discs.
This has an observable effect on the location of the CO snowline, which is located >200~au in most models for a 3~M$_{\odot}$ star, but is always within 80~au for 0.5~M$_{\odot}$ star. The chemical compositions calculated show that a well defined stellar mass and age range exists in which high C/O gas giants can form. In the case of the exoplanet HR8799b, our models show it must have formed before the star was 1~Myr old.

\end{abstract}




\section{Introduction}


The host star has a major influence on both the evolution of protoplanetary discs and on the properties of the planetary systems that they go on to form. Giant planets occur more frequently around more massive stars, specifically radial velocity surveys find a peak in giant planet occurrence rate for a host stellar mass of 1.9~M$_\odot$ \citep{Johnson2010, Reffert2014}. ALMA surveys of star forming regions show that the mass of solids in protoplanetary discs also scales with stellar mass \citep{Pascucci2016ARELATION, Mulders2018}. This trend complies with theoretical predictions; the core accretion scenario for planet formation predicts more giant planets to be formed around higher mass stars \citep{Ida2005TOWARDMASSES,Kennedy2008}. Furthermore planet population synthesis models by \citet{Alibert2011ExtrasolarMasses} show that scaling disc mass with the mass of the stellar host can explain the lack of high mass planets around 0.5~M$_\odot$ hosts and a lack of short separation planets around stars with mass greater than 1.5~M$_\odot$.

The central star determines the temperature structure within the disc it hosts. More massive, more luminous stars result in a warmer disc, which is expected to result in changes in the disc composition. 
Solids in the disc generally hold a greater relative abundance of oxygen compared to the gas which is instead relatively carbon-rich. Based on observations of gas and dust in discs, \citet{Oberg2011} take the ratio of carbon to oxygen, C/O, in both the gas and solid phase is dictated by the location of snowlines for key molecular carriers of oxygen and carbon. For example in their model the gas is most carbon-rich between the CO$_2$ and CO snowlines, because here the gas phase is dominated by CO, leading to C/O$\sim$1. 

An exoplanet's overall composition is inherited from the local disc at the time of formation when material is accreted rapidly. Modelling of the molecular composition of protoplanetary discs is thereby a useful tool for considering the composition of the building blocks of planet formation \citep[e.g.][]{Eistrup2016, Bosman2018,Cridland2019ConnectingAstrochemistry,Booth2019}.  Measurements of C/O and of metallicity (e.g. C/H, O/H) in exoplanets can be used to connect the composition of the planet with the location of its formation, and the relative amounts of oxygen-rich solids and carbon-rich gas this would have required. For example \citet{Brewer2017C/OLINE} investigate C/O in hot Jupiters and find that super-stellar C/O and sub-stellar O/H ratios suggest that some hot Jupiters form beyond the H$_2$O snowline and migrate inward. 
The enrichment of the atmospheres of giant exoplanets is aided by radial drift in the disc, which acts to carry solids to shorter radial separations from the star, doing so at a rate faster than depletion due to chemical processes \citep{Booth2017, Krijt2018TransportDrift, Booth2019}. The inward drift of pebbles is also an effective mechanism for transporting volatiles that are frozen onto grains towards the inner regions of the disc. This causes a plume of volatile gas to build up at snowline locations, which viscously spreads to enhance gas phase abundances of dominant molecular species within their own snowlines \citep{Stammler2017a,Booth2019}.
Exoplanet atmospheric C/O may differ from the exoplanets overall composition due to chemical effects within the atmosphere that we are not taking into account with this model \citep[for a detailed description of factors affecting atmospheric chemistry see review by][]{Madhusudhan2016ExoplanetaryHabitability}.

The formation of planets via gravitational instability on the other hand involves the large scale collapse of a mixture of gas and dust. The compositions of objects forming in this manner will likely reflect the stellar value \citep{Madhusudhan2014}. However, thermal history and internal sedimentation in these fragments can form objects with sub-stellar C/O \citep{Ilee2017}.

The temperature in the disc does not necessarily remain constant however, meaning the snowline locations are not fixed to a particular radial distance from the star. By inspecting a Hertzsprung Russel diagram or calculating the balance between gravitational collapse and nuclear reactions in a forming star \citep{Iben1965StellarSequence.}, the difference in luminosity evolution between young stars of different mass is clear \citep{Palla1993TheStars,Siess2000AnStars}. 
Low mass stars decrease in luminosity as they evolve towards the main sequence meaning there is a decreased irradiating flux upon the disc. Intermediate mass stars on the other hand, increase in effective temperature as they evolve towards the main sequence, but do not experience as large a decrease in luminosity as low mass stars, in fact after a few Myr their luminosity begins to increase again. 
Due to this contrasting evolution of low-mass stars compared to intermediate mass stars, \citet{Panic2017} infer different physical, and therefore chemical, histories in the discs around low- and intermediate-mass stars. 
So without the need to invoke any internal disc evolution, we can expect the temperature structure of the disc to evolve alongside the luminosity of the host star, introducing a temporal dependence for composition, in addition to the radial dependence from \citet{Oberg2011}. Every star must undergo this evolution, therefore every disc must experience the repercussions to some extent.

Furthermore the temperature structure of a disc also influences the mechanisms by which planet formation takes place. 
Disc temperature is important for the transport of solids in the disc because the speed of radial drift depends on disc temperature, v$_{\rm drift} \propto \Big( \frac{\rm H}{\rm r} \Big)^2 \propto$ T . It is also important for the composition of solids, as the snowlines dictate the presence of ices on the dust. In inner regions grain size is limited by fragmentation \citep{Birnstiel2012ADisks}. The fragmentation velocity is higher for grains possessing an icy mantle in comparison to ice-free grains such as those interior to snowlines from which volatiles have been released into the gas phase. A transition from mainly ice-free to icy grains results in a higher fragmentation velocity and therefore a sharp increase in the grain sizes outside of the H$_2$O snowline \citep{Gundlach2015THEPARTICLES}. 
Pebble accretion is a mechanism by which planetesimals can grow beyond $\sim$100s of km in size into cores of $\sim$10s M$_\oplus$ \citep{Johansen2010ProgradeEnvironment,Ormel2010TheDisks,Lambrechts2012RapidAccretion}. In combination with the streaming instability \citep{Youdin2005StreamingDisks,Johansen2007RapidDisks}, this offers a solution to the metre-sized barrier. It relies on the presence of dust particles of size $\geq$cm that are accreted as a result of passing through the Hill sphere of large planetesimals. If stellar evolution determines if midplane temperature profile, then there is potential to determine temporal constraints on when and where pebble accretion can operate in a disc, by virtue of whether or a reservoir of pebble sized dust particles can exist or is limited by fragmentation. 
 
In this paper we explore the impact that stellar evolution has on midplane conditions, in terms of the physical and chemical structure in the midplane of protoplanetary discs. 
We use self consistent temperature profiles calculated using stellar parameters from throughout the pre-main sequence evolution of stars across a range of stellar mass (spectral type). The relative abundance of key C and O carriers are calculated by applying the \citet{Oberg2011} chemistry model and considering the effects of radial drift, viscous evolution, diffusion, and the growth and fragmentation of dust. Gas mass loss from the disc has a significant effect on temperature, resulting in a less flared, cooler disc with a greater degree of dust settling \citep{Panic2017}. In order to isolate only the effects of stellar evolution, we do not evolve the mass of the disc, instead we provide snapshots of a range of disc conditions in which disc mass is high enough to allow giant planet formation.
We apply this approach in a case study of the system of HR8799, a F0 type star \citep{Gray2003CONTRIBUTIONSI.} with four directly imaged giant planets between which there are substantial variations in the atmospheric C/O retrieved from their observed spectra \citep{Lavie2017HELIOS-RETRIEVAL:Formation}. 

\begin{figure}
\centering \includegraphics[width=0.49\textwidth]{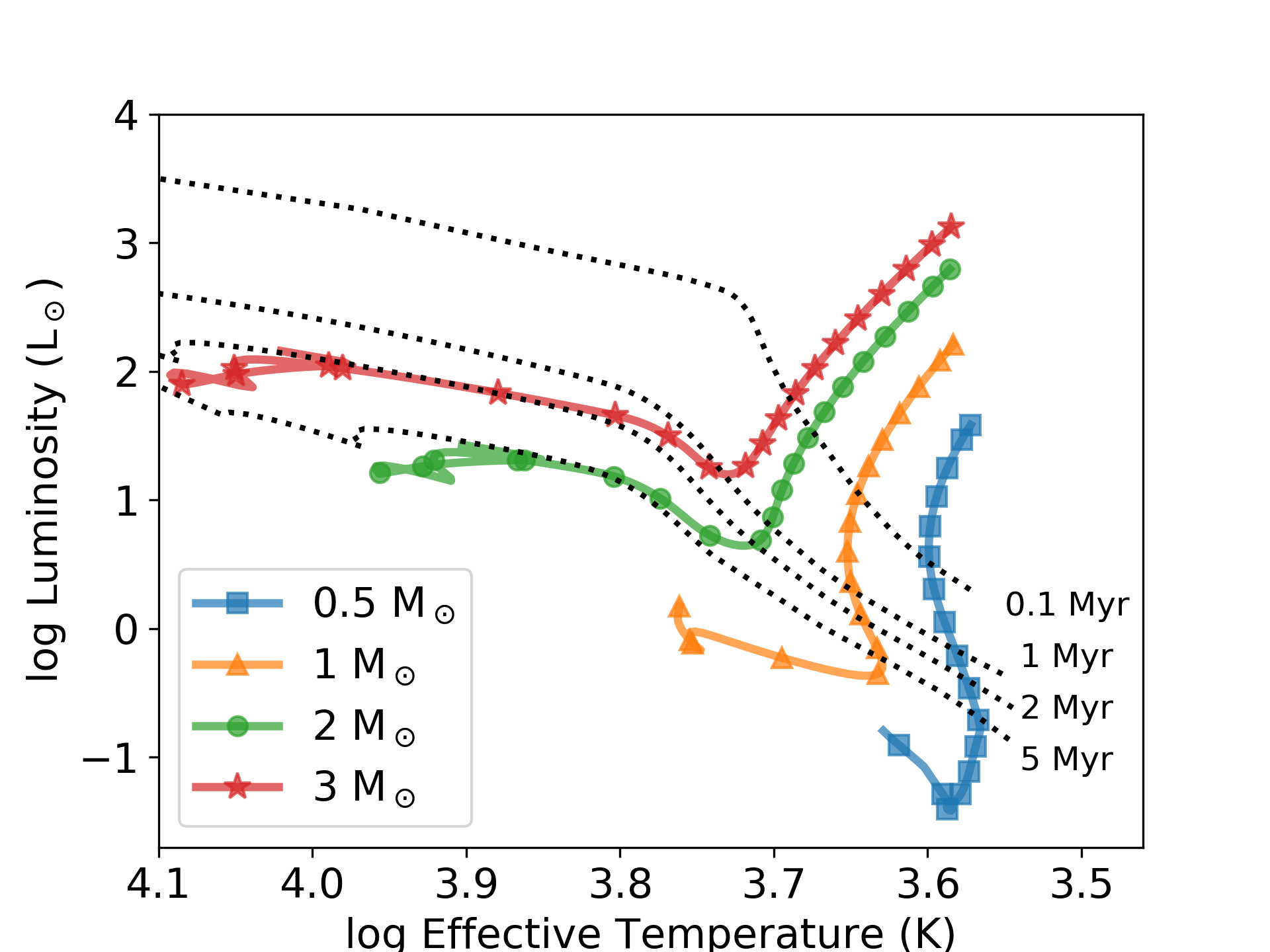}
\caption{Results of the stellar evolution models, following four stellar masses for 10~Myr. Dotted lines mark the isochrones for selected timesteps in the stellar evolution as labelled.}
\label{fig:hrd}
\end{figure}

\section{Methods}
\label{sec:methods}

\subsection{Physical Modelling}
\label{sec:phys_mod}
We combine three modelling codes to study the impact of stellar evolution on protoplanetary disc midplanes.
First, a grid of stellar parameters are calculated using the MESA code \citep[][]{Paxton2011MODULESMESA,Paxton2013ModulesStars} by evolving stars of mass 0.5, 1.0, 2.0, 3.0 M$_\odot$ for 10~Myrs. These masses were chosen to span the range over which there is a peak in planet occurrence rate in measured from radial velocities \citep{Reffert2014}.
Figure \ref{fig:hrd} plots this evolution on a Hertzsprung-Russel diagram, alongside MIST\footnote{http://waps.cfa.harvard.edu/MIST/} isochrones \citep{Dotter2016MESAISOCHRONES}.

The computed stellar parameters are used as input to radiative transfer models. We use the Monte Carlo radiative transfer code MCMax \citep{Min09} to iteratively solve for self-consistent temperature and vertical structure throughout the disc. The disc surface density is described by power law with an exponential tail \citep{Andrews2011RESOLVEDDISKS} consistent with the self-similar solution for a viscously evolving disc with a viscosity scaling as $\nu \propto R^\gamma$ \citep{Lynden-Bell1974TheVariables, Hartmann1998AccretionDisks}:

\begin{equation}
    \Sigma_{\rm gas} = \Sigma_{\rm c} ~ \Bigg( \frac{R}{R_{\rm c}} \Bigg)^{-\gamma} \rm ~ exp \Bigg[ - \Bigg( \frac{R}{R_{\rm c}} \Bigg)^{2-\gamma} \Bigg]~,
\end{equation}
where R$_{\rm c}$ is the critical radius and $\Sigma_{\rm c}$ the surface density at the critical radius. We adopt a disc mass of 0.1~M$_\odot$, ensuring the discs would be sufficiently massive to form multiple giant planets. We assume this large disc mass partly to reflect the higher mass expected in the early embedded stages of a disc when the planet formation process begins \citep{Tobin2020TheDisks, Zhang2020RapidStage}, and because observational disc masses are expected to be significantly underestimated \citep{Miotello2018,Booth2019HD132Mass}. The dust in the disc model is comprised of astrosilicate grains with an MRN size distribution, $ dn/da \propto a^{-3.5} $ \citep{Mathis1977}. Parameters adopted for the MCMax models are summarised in Table~\ref{tab:RT_params}.

\begin{table} 
\caption{Disc parameters adopted in radiative transfer modelling}             
\label{tab:RT_params}      
\centering                          
\begin{tabular}{l l c}        
\hline\hline                 
Parameter & & Value(s) \\    
\hline                        
    R$_{\rm in}$ & Inner radius & 0.1~\rm{au}     \\
    $\gamma$ & Initial $\Sigma(R)$ slope & 1.0 \\
    R$_{\rm c}$ & Critical radius & 200~\rm{au}            \\
    a & Grain sizes & 0.01~$\mu$\rm{m} - 10~\rm{cm}  \\
    $\alpha$ & Turbulent viscosity   & 10$^{-3}$             \\
    g/d & Gas-to-dust mass ratio & 100\\
    M$_{\rm d}$ & Dust mass & 10$^{-3}$ M$_\odot$\\
\hline                                   
\end{tabular}
\end{table}

We model the disc evolution as in \citet{Booth2017} in order to calculate transport within the disc by taking into account viscous evolution, radial drift of solids and diffusion, as well as the growth and fragmentation of dust grains following the method of \citet{Birnstiel2012ADisks}. The midplane dust and gas are allowed to evolve for 10~Myr, the temperature profile is updated throughout this process with the results of the radiative transfer modelling. Radiative transfer models were generated at steps for 0.01, 0.1, 1.0, 2.0, 3.0, 5.0 and 10.0~Myr, in between these points we interpolate between the temperature profiles for adjacent timesteps, assuming a linear evolution with time. As a result of inward transport of dust and gas, some material is accreted onto the central star. 

Evolution of disc mass was not included in the radiative transfer calculations. The extent to which $\Sigma$(R) is altered by drift and viscous evolution depends primarily upon the size of the disc (determined by R$_c$ in our modelling) and on the turbulent viscosity $\alpha$, and affects inner regions of the disc the most. We take R$_c$=200~au and $\alpha=10^{-3}$, for these parameters $\Sigma$(R) is not depleted in any model by more than 30\% up to 5~Myr, by 10~Myr $\Sigma$(<50au) been depleted by $\sim$ 50\%. The removal of gas mass will decrease temperature in the disc as demonstrated in \citet{Panic2017}, so our temperature profiles can be taken as warm upper limits. 

The dust evolution we calculate over a 10~Myr timescale will have implications for the photo-surface of the disc that collects starlight and heats the disc. The most important factors in this regard however are the minimum grain size (as it is the small grains that reside in the upper layers of the disc that absorb the starlight) and the gas mass \citep{Panic2017}. Our choice of a small a$_{\rm min}$ and large gas mass will maximise the heating in our disc models, and so the temperatures we calculate offer an upper limit in which heating is the most effective.

The justification for calculating the physical evolution of the midplane separately from the radiative transfer modelling is that a grid of models can be created that represent massive discs such as TW Hydra \citep{Bergin2013} and HD163296 \citep{Booth2019HD132Mass}, i.e. massive protoplanetary discs with ages beyond that of typical near-IR disc lifetimes.

\subsection{Chemical model}
\label{sec:chem_mod}

We investigate the effect of temperature evolution as a result of changing stellar luminosity on the disc composition using the code of \citet{Booth2017}. The code treats the adsorption and desorption of the main C and O carriers along with their transport. We adopt the same key C and O carrying molecules and their abundances in the midplane from \citet{Oberg2011} as given in Table \ref{tab:Oberg_chem}, also implemented as case 2 in \citet{Madhusudhan2014,Madhusudhan2017, Booth2017}. Solar initial abundances are adopted \citep{Asplund2009TheSun}, following \citet[][]{Konopacky2013DetectionAtmosphere}. 

Transport is included in the gas phase and on the surface of grains due to radial drift. Equilibrium adsorption is treated following \citet{Hollenbach2009WaterClouds} with equilibrium assumed, and is dependent upon the size, number density and number of binding sites on the dust \citep{Hasegawa1992ModelsMoleculesb,Visser2009}. The thermal desorption rate per grain is a function of the surface density of binding sites parameter. We use 0.1 micron grains in computing the adsorption / desorption equilibrium, under the assumption that small grains dominate the surface area. For this grain size all of the binding sites are occupied, resulting in snow line locations that are not sensitive to grain size. This gives results consistent with models that use an average grain size (by area) \citet[][]{Booth2019}.  

More detailed chemical reactions are not required as we assume transport processes to be dominant due to efficient radial drift \citep{Booth2019}, except maybe on time-scales of several Myr where grain chemistry can be important \citep{Krijt2020COProcessing}. 

The evolving stellar evolution sets the radial temperature profile as described above, which in turn determines the position of snowlines for the key molecules in the disc model.  
We note that a number of factors can shift snowlines slightly in the disc \citep{Lecar2006OnDisk,Kennedy2008PlanetPlanets,Fayolle2016N2Ice}. The main purpose of this work however is to investigate global trends with varying stellar mass and age, so for simplicity we fix the binding energies to those given in Table \ref{tab:Oberg_chem} and the CO, CO$_2$ and H$_2$O snowline positions to characteristic disc temperatures of 20, 47 and 135~K respectively, following \citet{Oberg2011}.

\begin{table} 
\caption{Binding energies, presented as temperatures, and volume mixing ratios of the key chemical species included in our chemical model. Binding energies are from \citet{Piso2016THEDISKS} and volume mixing ratios follow \citet{Oberg2011}.}
\begin{tabular}{ccc}
\hline \hline
Species              & T$_{\rm bind}$ (K) & X/H \\ \hline
CO                   & 834   & 0.65 $\times$ C/H  \\
CO$_2$               & 2000  & 0.15 $\times$ C/H \\
H$_2$O               & 5800  & O/H - ( 3 $\times \frac{\rm Si}{\rm H}$ + $\frac{\rm CO}{\rm H}$ + 2 $\times$ $\frac{\rm CO_2}{\rm H}$ ) \\
Carbon grains        & n/a & 0.2 $\times$ C/H  \\
Silicates            & n/a & Si/H  \\
\hline
\multicolumn{1}{l}{} & \multicolumn{1}{l}{}    & \multicolumn{1}{r}{}                           
\end{tabular}

\label{tab:Oberg_chem}
\end{table}

\section{Results}
\label{sec:results}

\subsection{Midplane temperature}
\label{sec:res_RT}
\begin{figure} 
\centering \includegraphics[width=0.49\textwidth]{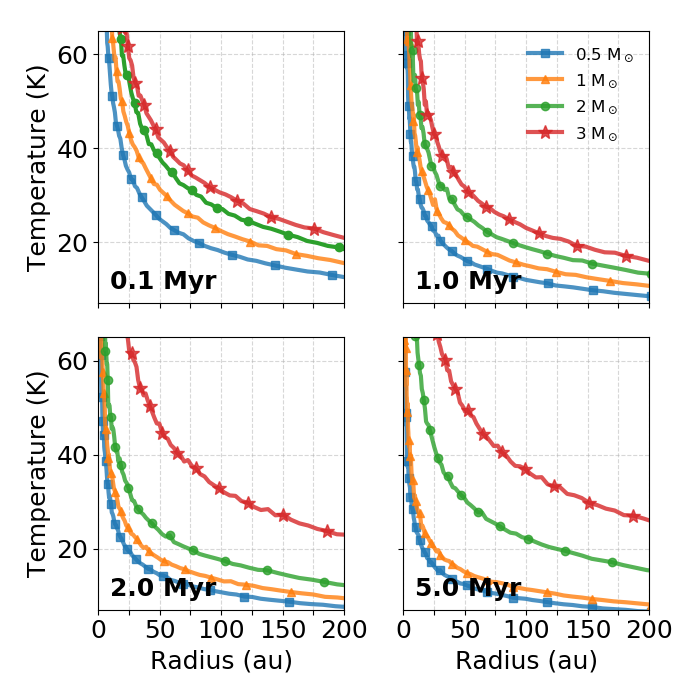}
\caption{Snapshots of the midplane temperature as a function of radius around stars of 0.5, 1, 2 and 3~M$_\odot$ at different stellar ages. Each panel represents a different snapshot in time, each line represents a disc around a different star as labelled in the key.  }
\label{fig:midplaneT_ageevol}
\end{figure}

Figure \ref{fig:midplaneT_ageevol} shows a selection of midplane temperature profiles resulting from our radiative transfer calculations. Each panel shows the midplane temperature profile in the disc that results from the stellar parameters at different timesteps, as labelled in the bottom left corner of each plot. In each panel we plot the results of discs around stars with mass 0.5 - 3 M$_\odot$. 

In the top left panel of Figure \ref{fig:midplaneT_ageevol} the temperature profiles for different M$_*$ are relatively similar to each other for a stellar age of 0.1~Myr, but begin to diverge at later time steps. At 0.1 Myr, at 25~au the disc around the 3.0~M$_\odot$ star is $\approx$70\% hotter than that around the 0.5~M$_\odot$ star. At 5~Myr, the 3.0~M$_\odot$ case is $\approx$440\% hotter than the 0.5~M$_\odot$ one. This divergence of midplane temperatures follows from the differing evolutionary paths taken by the host stars. In Figure \ref{fig:hrd} the difference in luminosity between the 0.5 and 3~M$_\odot$ stars is also significantly larger at 5~Myr than it was at 0.1~Myr. 

\begin{figure} 
\centering \includegraphics[width=0.49\textwidth]{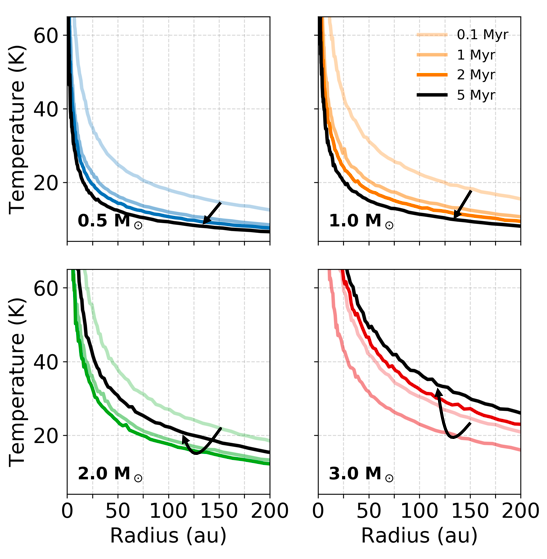}
\caption{Midplane temperature as a function of radius at different times for different M$_*$. Each panel shows a model with a different central star (labelled in the bottom left corner) and each curve represents a different time step, with the lightest colour representing the earliest time. The arrow on each plot links the 4 curves in chronological order.}
\label{fig:midplaneT_stellarmass}
\end{figure}

The effect of this dichotomy in evolution is shown clearly in Figure \ref{fig:midplaneT_stellarmass}, where each panel contains models with the same stellar mass plotting temperature curves for different time steps. Whilst the discs around low-mass/solar-like stars in the top panels cool with age (see the black arrows in Figure \ref{fig:midplaneT_stellarmass}), the discs around intermediate mass stars experience an increase in midplane temperature due to the increasing luminosity of the host star meaning they are relatively warm at late times. For example, in the 3~M$_\odot$ case (bottom right panel Figure \ref{fig:midplaneT_stellarmass}), at a separation of 25~au the disc is in fact 38~K warmer at 5~Myr than it was at 1~Myr. 

These models quantitatively demonstrate that not only are the discs around intermediate mass stars warmer than around their low-mass counterparts, but that they also remain warmer for longer corresponding to the evolution of the host star's luminosity. 

\subsection{Snowline locations}
\label{sec:snow_loc}
\begin{figure} 
\centering \includegraphics[width=0.49\textwidth]{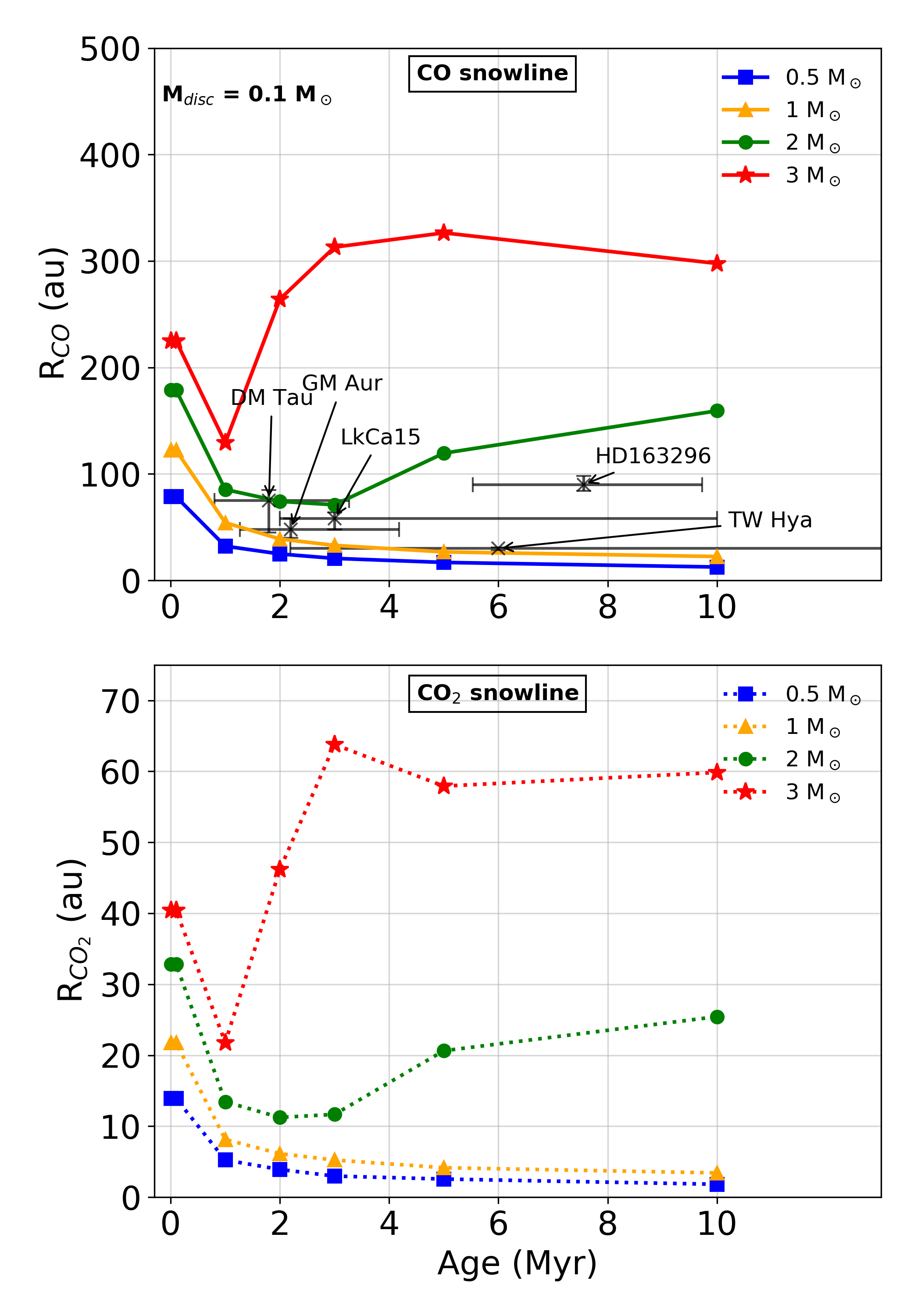}
\caption{Position of the CO (T=20K, top panel) and CO$_2$ (T=47K, bottom panel) snowlines in the disc midplane around stars of different mass over a range of time steps. Each model has a total disc mass of 0.1~M$_\odot$. Crosses show the positions of N$^2$H$^{+}$ rings, tracing the CO snowline location of protoplanetary discs GM Aur, LkCa15, DM Tau, HD163296 and TW Hydra. The disc temperatures for the observed snowlines are: GM Aur 24-28~K, LkCa15 21-25~K, DM Tau 13-18~K, HD163296 25~K and TW Hydra 17~K. }
\label{fig:Rco}
\end{figure}

A direct consequence of the variation in midplane temperature profile is variation in the location of snowlines within the disc. Figure \ref{fig:Rco} shows the CO and CO$_2$ midplane snowline locations, assumed to be at T=20~K and T=47~K respectively, around different masses of star. 
Disc temperature at the snowline of a species can vary somewhat depending on the ices that exist on the grain. Laboratory experiments have shown CO desorption energies to vary by $\ge$50\% depending on the water coverage and structure of the ice in which it is adsorbed. The range in binding energies corresponds to a range in snowline temperatures, for example for an assumed midplane density (n$_H=10^{10} \rm cm^{-3}$) and CO abundance relative to n$_H (5\times10^{-5})$ \citet{Qi2019ProbingEmission} point out that a desorption energy for CO adsorbed to compact H$_2$O ices (E$_{\rm des}$=1300~K) corresponds to a CO snowline temperature of $\sim$ 32~K, whereas considering only pure ices (E$_{\rm des}$=870~K) corresponds to a snowline temperature of $\sim$21~K. As mentioned previously we adopt constant T in order to compare global trends across stellar mass and age.

The discs around 0.5 and 1~M$_\odot$ stars are at their coolest at later times, and so the snowlines are at shorter separation from the host star. Figure \ref{fig:Rco} shows that at 2~Myr both discs have CO snowlines that are within 50~au. Conversely the discs around intermediate mass stars begin to warm up at later times, pushing the CO snowline further out into the disc. For example in Figure \ref{fig:Rco} the 3~M$_\odot$ star CO snowline is found at a distance of over 250~au from the host star after 2~Myr, meaning that CO at large orbital radii will remain in the gas phase for a large proportion of the disc lifetime. 

In Figure \ref{fig:Rco}, derived locations of CO snowlines in observed systems have been plotted alongside results from our models. In each case the snowline location has been derived from observations of rings of N$_2$H$^+$ by \citet{Qi2013,Qi2015CHEMICALDISK,Qi2019ProbingEmission}.  N$_2$H$^+$ is a reactive ion abundant where CO is frozen out and therefore a useful CO snowline tracer. 

In Figure \ref{fig:Rco} there is a broad agreement with the expected trends especially in the older systems HD163296 and TW Hydra. 
HD163296 is a 1.9~M$_\odot$ star that has a midplane CO snowline in its disc at around 90$\pm$8~au \citep{Fairlamb2015,Qi2015CHEMICALDISK}, whereas TW Hydra has a stellar mass of 0.8 and a CO snowline much closer in at 21$\pm$1~au. Despite both discs being around the same age, the disc of HD163296 is warmer than TW Hydra supporting the results of our radiative transfer modelling. The CO snowline prediction for a 2~M$_\odot$ star in Figure \ref{fig:Rco} is further out than the  location fitted from observations of N$_2$H$^+$ in HD163296, in part due to the fact that the observed N$_2$H$^+$ ring was found at 25~K rather than 20~K \citep{Qi2015CHEMICALDISK}, and also due to the assumption of a continuous disc, whilst the system famously hosts multiple large gaps. The observed location of the TW Hya CO snowline is further out in the disc than our estimate, which may be due to its low mass in comparison to that which we assume in the model; modelling of HD observations find a disc gas mass $\geq 0.06$. 
At early times in contrast it is difficult to distinguish midplane temperatures by stellar mass because stellar luminosity is at a minimum and the midplane disc temperatures are at their most similar. The three youngest discs are around stars of lower mass GM Aur (1.3~M$_\odot$), LkCa15 (1.0~M$_\odot$) and DM Tau (0.5~M$_\odot$). In Figure \ref{fig:Rco} the location of the CO snowline in GM Aur agrees with the prediction of the disc models using stars of similar mass. The CO snowlines in LkCa and DM Tau are slightly further out in the disc, by $\approx $20~au, than our models predict. This is likely due to the depleted nature of their inner discs; LkCa 15 has an inner cavity of $\sim$45~au \citep{Pietu2006ResolvingWavelengths,Jin2019NewALMA} and DM Tau has two thin dust rings at 4~au and 25~au with a more extended outer disc from $\approx$75~au \citep{Kudo2018ATau}. In both these systems, the disc at 10s of au is therefore likely to be much warmer than a similar location in our assumed disc model as described in Table \ref{tab:RT_params}.

While the providing a first order validation, we note that using N$_2$H$^+$ as a tracer of the CO snowline can be complex as it is not completely clear where the emission originates from, considering vertical temperature structure is crucial. Modelling by \citet{vantHoff2017RobustnessSnowline} demonstrates that N$_2$H$^+$ emission peaks exterior to the CO snowline (by at least $\sim$5~au in their models of a TW Hya-like disc). Furthermore N$_2$H$^+$ can become increasingly abundant in the surface layer to a degree that depends on vertical temperature structure and CO abundance. In discs where dust has settled this results in peaks that are closer to the star than observed and further chemical modelling should be employed in order to derive a CO snowline location \citep{vantHoff2017RobustnessSnowline}.

N$_2$H$^+$ has not yet been detected in embedded disc at early stages. In modelling of a forming protoplanetary disc around a solar mass star, in a viscously spreading disc  \citet{Drozdovskaya2016CometaryMidplanes} find a CO snowline at $\approx$0.25~Myr beyond 50~au, consistent with the closest model of ours which has a CO snowline at 70~au for the same age. Disc formation processes influence their disc strongly however, the infall dominated model finds a CO snowline closer in at 23~au.

\subsection{Disc Composition}
\label{sec:comp}

We now explore the effects of the CO snowline locations, derived in Section \ref{sec:snow_loc}, on the composition of gas and ice in the disc midplanes. Just as our models place the upper limits on midplane temperatures, so the CO snowline locations represent the furthest possible distance where the CO snowline may be found at the given age/mass.

\begin{figure} 
    \centering
    \includegraphics[width=\linewidth]{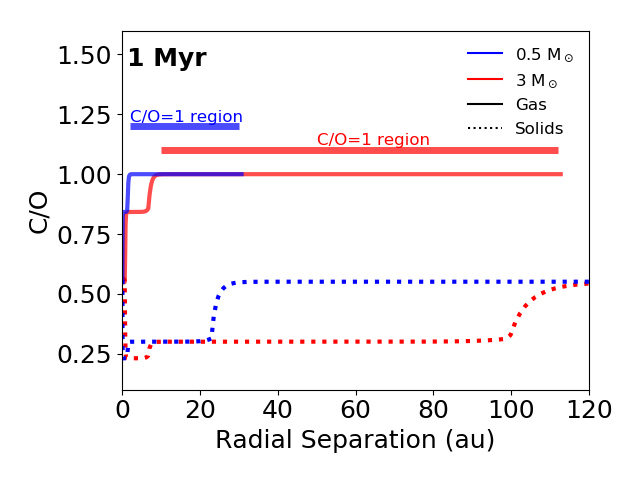}
    \caption{C/O in the gas (solid line) and solid phase (dotted line) for models at 1~Myr with blue lines showing the model with a 0.5~M$_\odot$ host star, and red lines showing a 3~M$_\odot$ host star. No drift or transport are included in this instance. The gas phase line is plotted up to the location of the CO snowline in the midplane, taken to be at T=20K. Horizontal lines above the C/O curves denote the region for which gas phase C/O=1 for each of the models and are colour coded accordingly.}
    \label{fig:simple_comp_spectype}
\end{figure}

Figure \ref{fig:simple_comp_spectype} plots C/O as a function of radius for the two extremes of our stellar mass range, the 0.5 and 3~M$_\odot$ cases, calculated without including any drift or transport within the disc, meaning the only difference should be due to the temperature profile. Each reproduces the same step function as seen in \citet{Oberg2011}, presented here on a linear radial scale. The warmer disc around the 3~M$_\odot$ star pushes both the CO and CO$_2$ snowlines further out in the disc, leading to a more extended region of maximum carbon enrichment in the gas phase (in our model this is where C/O=1) in comparison to the disc around the 0.5~M$_\odot$ star. Outwards of $\sim$30~au in Figure \ref{fig:simple_comp_spectype}, the solids in the 0.5~M$_\odot$ case are much more carbon enriched because all main carbon carriers are frozen out onto the grains. In the warmer 3~M$_\odot$ case, the CO snowline is at $\approx$115~au in the disc and so more carbon is in the gas phase, leaving solids with a lower value of C/O $\approx$ 0.3 . We note that intermediate mass stars produce more FUV photons than their lower mass counterparts, which could result in increased production of CO$_2$ in the disc in the inner (<10~au) parts of the disc \citep{Walsh2014a,Drozdovskaya2016CometaryMidplanes}. 

\begin{figure} 
    \centering
    \includegraphics[width=\linewidth]{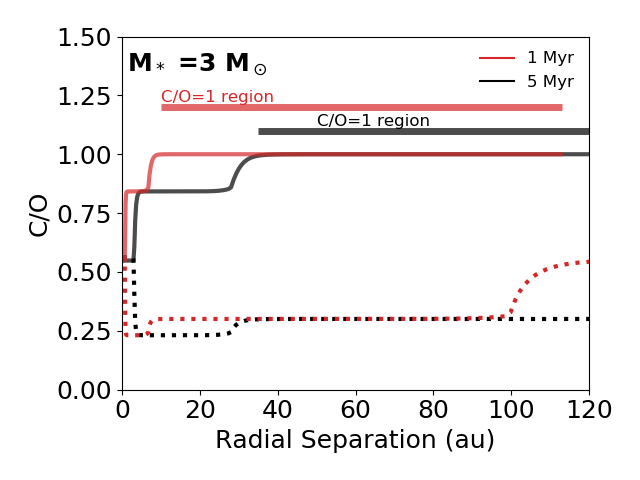}
    \caption{C/O in the gas (solid line) and solid phase (dotted line) for models with a 3~M$_\odot$ host star and no drift or transport included. The red lines show the model with stellar parameters at 1~Myr, the black lines show the model at 5~Myr. The gas phase line is plotted until the location of the CO snowline in the midplane, taken to be at T=20K. Horizontal lines above the curves denote the region for which gas phase C/O=1 for each of the models and are colour coded accordingly. }
    \label{fig:simple_comp_evo}
\end{figure}

Heating of the disc depends not only on the mass of the star, but the stage of its evolution as well. In Figure \ref{fig:simple_comp_evo} we plot the same 1~Myr, 3~M$_\odot$ profile as in Figure \ref{fig:simple_comp_spectype}, but now compared to a model using stellar parameters at 5~Myr when the intermediate mass star's luminosity is greater. In the 5~Myr case the CO$_2$ snowline is further out than in the 1~Myr case. As shown in Figure \ref{fig:simple_comp_evo}, this results in a lower C/O ratio for the oxygen enriched solids up to 30~au in the 5~Myr case (dotted black line) compared to the 1~Myr case (red dotted line). CO remains in the gas phase for the vast majority of the disc extent, with the CO snowline found at over 400~au at 5~Myr. Figure \ref{fig:simple_comp_evo} therefore makes it clear that there is not only a radial dependence to be considered, but also a temporal one. The CO and CO$_2$ snowline locations will evolve with the disc temperature profile resulting in changes to disc composition.

The H$_2$O snowline is always at very short separation from the central star, where inner regions are viscously heated rather than radiatively, and so we do not expect the H$_2$O snowline to be affected in a similar manner \citep[see e.g. ][for a study on transport and redistribution of water in discs.]{Ciesla2006TheDisk}

\section{Discussion}

\subsection{Application to HR8799}
\label{sec:HR8799}
The C/O of disc material in the gas and solid phases depends on both location within the disc and the age of the stellar host in our models. We now explore to what extent these results can explain variation in the atmospheric C/O of giant planets that have all formed within the same system.

HR8799 is an F0 type star \citep{Gray2003CONTRIBUTIONSI.} that hosts four directly imaged giant planets in its system \citep{Marois2008,Marois2010}. The wide orbits and large masses of the planets have proved difficult to model with current theories of planet formation. Through spectroscopic analysis \citet{Sadakane2006218396} find the abundances of C and O in HR8799 to be consistent with solar values; log$_{10}$C/H= -3.57 and  log$_{10}$O/H= -3.31 \citep{Asplund2009TheSun}. In this section we modify the disc modelling presented in Section \ref{sec:results} to be match to the stellar mass of HR8799, including stellar evolution, disc transport and the chemical model following \citet{Oberg2011}. We then compare the composition of the model disc with the C/O determined for each of the four exoplanets from their observed spectra by \citet{Lavie2017HELIOS-RETRIEVAL:Formation} using the atmospheric retrieval code HELIOS. Properties of the four exoplanets are summarised in Table \ref{tab:HR8799}. 

\begin{table}
\centering
\begin{tabular}{llllll}
\hline \hline
Planet &  Mass $^a$ & R$_{\rm pl}$ $^{a,b}$& C/H$^{c}$ & O/H$^{c}$& C/O$^{c}$  \\ \hline
\multicolumn{1}{l|}{HR8799b} & 7~M$_{\rm J}$ & 68~au & 1.46 & 1.24 & 0.92\\
\multicolumn{1}{l|}{HR8799c} & 7~M$_{\rm J}$ & 38~au & 0.84 & 0.84 & 0.55 \\
\multicolumn{1}{l|}{HR8799d} & 7~M$_{\rm J}$ & 24~au  & -13.0 & 0.73& $>10^{-13}$\\
\multicolumn{1}{l|}{HR8799e} & 5~M$_{\rm J}$ & 15~au   & -8.4 & 0.28 & $>10^{-9}$\\
\hline
\end{tabular}
\caption{Key properties of the giant exoplanets in the HR8799 system. The retrieved carbon abundance is very low in planets `d' and `e', and so the C/O represents a lower limit. C/H and O/H values are given relative to stellar values in the form: log$_{10}$X$_{pl}$ - log$_{10}$X$_*$, where X is the abundance relative to hydrogen.  References a) \citet{Marois2010}, b) \citet{Marois2008}, c) \citet{Lavie2017HELIOS-RETRIEVAL:Formation}. }
\label{tab:HR8799}
\end{table}

\begin{figure*}
    \centering
    \includegraphics[width=1\linewidth]{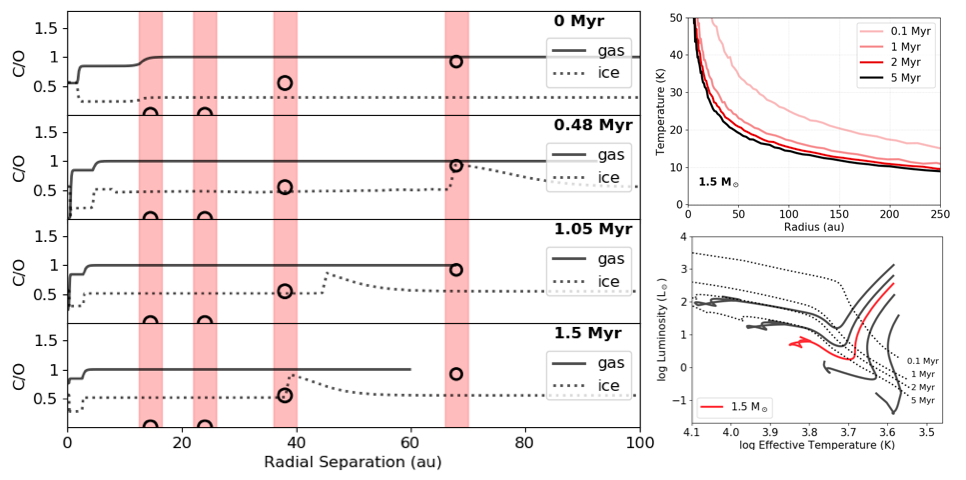}
    \caption{The three elements of the HR8799 disc model. \textbf{Left hand side:} Ratio of carbon to oxygen abundance in the gas phase (solid line) and solid phase (dotted line) as a function of radius. Shaded areas mark the location of the giant planets in the HR8799 system, black circles are plotted at the corresponding C/O retrieved for their atmosphere. Each panel shows a different time step in the evolution of our model as is labelled in the top right corner of each panel. Timesteps are chosen to reflect key points as referred to in the text. The gas phase C/O line is plotted up until the CO snowline in the midplane, taken to be at 20K. \textbf{Right hand side:} On the top is the evolution of T(R) in our model for a 1.5~M$_\odot$ star. On the bottom is an HR diagram comparing the pre-MS evolution of a 1.5~M$_\odot$ star to the evolutionary tracks in Figure \ref{fig:hrd}.}
    \label{fig:HR8799}
\end{figure*}

We adopt the disc properties as described in Table \ref{tab:RT_params} with a star of mass 1.5~M$_\odot$ and we explore the conditions in the disc models at ages up to 10~Myr. Figure \ref{fig:HR8799} shows snapshots taken from this series of models that illustrate selected key milestones to be discussed in the text. Assuming that the atmospheric C/O listed in Table \ref{tab:HR8799}
is comprised of a mixture of rapidly accreted local gas and dust, we now analyse whether evolving disc composition as a result of stellar luminosity evolution can put temporal constraints on planet formation in the HR8799 system. We do not speculate on the formation of planets `d' and `e', the C/O values from \citet{Lavie2017HELIOS-RETRIEVAL:Formation} are very low, this is mainly due to the fact that the retrieved probability density functions for C/H in these planets do not constrain the values well. GRAVITY has detected CO in the atmosphere of HR8799e \citep{Lacour2019FirstE}. Future observations will help to determine the composition of these exoplanetary atmospheres.

To aid with this discussion we calculate the relative mixing proportions of gas and dust that would be required to achieve the C/O for each exoplanet atmosphere as given in Table \ref{tab:HR8799}. By assuming that the atmospheric C/O is inherited from local disc material, we can write the C/O of a giant planet as a mixture of local gas and dust,

\begin{equation}
     \rm C/O_{planet} = f~C/O_{gas} + (1-f)~C/O_{solid},
 \label{eqn:gas_frac}
\end{equation}
where $f$ is the fraction of material that is accreted to the planet that was originally in the gas phase in the disc. We assumed solar abundances in our modelling, but note that the initial disc composition is unlikely to be a direct inheritance from the initial cloud due to chemical evolution during the disc formation process \citep{Visser2009TheIces,Visser2011TheSpecies,Harsono2013EvolutionFormation,Drozdovskaya2016CometaryMidplanes}. 

In Figure \ref{fig:HR8799} we present selected snapshots from our modelling that show key time steps. On the left hand side are four panels showing C/O radial profiles for the HR8799 system. The black lines give the C/O in the solids and gas as labelled in the legend, the coloured bars give the orbital radii of the 4 known giant planets, with a circle plotted for each planet at a y value corresponding to the atmospheric C/O found by \citep{Lavie2017HELIOS-RETRIEVAL:Formation} for each planet. On the right hand side the midplane temperature profiles are given for different stages of the star's luminosity evolution (which can be compared to those in Figure \ref{fig:midplaneT_stellarmass}). The path of the 1.5~M$_\odot$ star across the HR diagram as calculated by our MESA modelling is shown in the bottom panel on the right hand side, with the tracks of the 0.5, 1, 2 and 3 M$_\odot$ stars also plotted in grey. 

In Figure \ref{fig:HR8799}, after 1.5~Myr, the evolution slows and the radial profile of C/O in gas and dust is well defined. From these key epochs C/O$_{\rm gas}$ and C/O$_{\rm solid}$ can be measured and used in Equation \ref{eqn:gas_frac}.

\subsubsection{HR8799b}

The main constraint our modelling puts on the formation of carbon-rich exoplanet HR8799b - the outermost planet in Figure \ref{fig:HR8799} - is a result of the displacement of the furthest radial location where the CO snowline may be located at different ages, as given by our models. At a radial separation of 68~au HR8799 is interior to the CO snowline position in all models up to the 1.05~Myr one, which is shown in Figure \ref{fig:HR8799}. Beyond the CO snowline, no major C or O carriers are in the gas phase. This means that rapid gas accretion in this particular region during planet formation will have no affect on the C/O ratio of any forming planet as the gas is mainly H$_2$, but it will affect the C/H ratio. HR8799b therefore must have accreted the vast majority of its carbon and oxygen while still inside the C/O=1 region (inside the CO snowline). 
Applying Equation \ref{eqn:gas_frac} to the orbital radius of HR8799b, $\geq80\%$ of the planet's carbon and oxygen must be from the gas phase in order to reproduce the atmospheric C/O=0.92 for the majority of the disc's evolution. Full calculations of $f$ for HR8799b and HR8799c are given in the Appendix.

The point in time when the CO snowline is located inside the planet's orbital radius depends on gas mass in the disc, as it is the amount of gas which determines the degree of flaring in the disc and its subsequent heating \citep{Panic2017}. Disc gas mass will decrease with time as a result of photoevaporation and accretion onto the star, resulting in a lower midplane temperature and a shorter time to form the planet. Furthermore if the planet experienced migration through the disc it would have accreted gas starting at a greater radial separation from the star where there is even less time before the CO snowline crosses it position. A time constraint for formation of $\sim$1~Myr is derived from our models by considering the time at which the CO snowline moves inside the planet's location in the radiative transfer models. This represents an upper temporal limit on the accretion of C and O from the disc, pointing towards an early formation time for HR8799b. 
In certain circumstances runaway migration can also lead to outward motion \citep{Masset2003RUNAWAYJUPITERS} which would leave the planet beyond the CO snowline once more. For example if a planet is located close to a sharp disc edge on the outside of a disc cavity \citep{Artymowicz2004DEBRISPlanets}. If this were to the the case for HR8799b, it would become very difficult to also explain the planets interior to it.  
Scattering events between giant planets can theoretically re-order the configuration of a planetary system, similarly to the Nice model for the Solar System \citep{Gomes2005OriginPlanets}. 
However, the compact and apparently stable orbits of the HR8799 planets suggests that scattering did not occur. While scattering followed by re-circulariziation of the orbits due to planet-disc interaction may be possible, it was rare in the population synthesize models of \citet{Forgan2018TowardsInteractions}.
If scattering events occur very early then interactions of the perturbed planets with the disc 
\citet{Fabrycky2010STABILITYMASSES} suggest the stability in the orbits of the system require mean-motion resonance between them, which is more likely to occur through migration into a stable state, rather than a chaotic scattering event.

\subsubsection{HR8799c}

HR8799c is found at an orbital radius of 38~au and so it remains inside the snowline in all our models, meaning it can always accrete gas containing carbon and oxygen carriers that will alter the C/O ratio of the planet. Table \ref{tab:HR8799} shows that HR8799c has C/O comparable to that of the central star which has C/O=0.56 \citep{Sadakane2006218396}, i.e. one that is much lower than the C/O in the gas phase of the disc models.

There is not much change in the composition structure after the 1.5~Myr model (bottom of Figure \ref{fig:HR8799}) as the stellar luminosity, and therefore radial temperature profile in the disc, does not significantly vary before the star joins the main sequence. 

Unlike HR8799b, the orbital radius of HR8799c means its enrichment cannot be constrained by CO snowline location, as it is found inside the CO snowline in all our models and also outside of the CO$_2$ snowline in all models. This means that local C/O in the gas and solids remains relatively stable as a function of time. This makes it more difficult to pinpoint a specific epoch for formation for a given exoplanetary atmospheric C/O. 

We can however exclude formation in the models between 1.5~Myr and 3~Myr, where solids are particularly high in CO, leading to a high C/O. 
This local increase in carbon-rich material is due to a build up in the outer disc at earlier times when the CO snowline was at greater radial separation. The build up occurs because grains drift faster than the volatiles diffuse, this produces local abundance increases of molecules near their snow lines \citep{Booth2017}. In our models the snowline moves inward over time, and the size of build up increases as surface density of the disc increases towards the star. CO-rich pebbles are transported inwards through radial drift after the CO snowline has rapidly moved inward due to the initial rapid decrease in host star luminosity, which occurs much quicker than radial drift timescales. 

The approximately stellar C/O of HR8799c is impossible to achieve when the carbon-rich material from the outer disc reaches the planet's position (See Figure \ref{fig:f_calc}). Using Equation \ref{eqn:gas_frac} to calculate the relative mixing of gas and ice required from the values of C/H and O/H at R=38~au throughout disc evolution, over 90\% of C and O must have originated from gas content in our models rather than solids. The solids that are accreted are highly rich in O, and reduce the overall C/O of the atmosphere. In order to achieve the approximately solar C/O of HR8799c in a region of the disc with carbon-enriched gas, there must have been substantial oxygen enrichment of the atmosphere through the accretion of solids. In our model, radial drift and the subsequent inward transport of volatiles from the outer disc are insufficient to achieve this. This indicates that the transport of CO rich material from the outer disc plays an important role in setting the composition of the inner disc, and that this impact is dependent upon time. Transport times for any individual disc will depend upon turbulent viscosity $\alpha$, and the surface density profile in the disc which in real systems may be perturbed by the opening of gaps or perhaps by infall of material from the wider envelope at earlier times.

A stellar C/O is seen in a number of hot Jupiters \citep[e.g. 8 out of 9 planets in the sample of][]{Line2014ARATIOS}. But unlike the hot Jupiters, HR8799c is on a relatively wide orbit. Changes in C/O of an exoplanet's atmosphere can also occur as part of the formation process. If planet formation occurs via gravitational instability, the C/O of a fragment can vary due to the sequestration of volatiles contained within dust that settles towards the fragment core \citep{Ilee2017} or due to the accretion of planetesimals \citep{Madhusudhan2014}. 
\citet{Kratter2010} identify a set of criteria to be met in order for any of the planets in HR8799 to have formed via GI. For their proposed scenario, the authors require a formation around the end of the Class I stage and beyond 40-70~au, with temperatures colder than typical discs. These criteria are consistent with our constraints of an early formation in the outer disc for HR8799b and the cooling of disc midplane as a result of stellar evolution. On the other hand if young discs are warm, as suggested by modelling \citep[][and our Figure \ref{fig:midplaneT_stellarmass}]{Harsono2015VolatileProtostars} and by recent ALMA observations of embedded class I discs in Taurus  \citep{vantHoff2020TemperatureWarm}, this may act to inhibit gravitational instability at early stages.

\subsection{Implications for planet formation}
Our radiative transfer models quantify the impact of stellar evolution on altering the temperature structure in the midplane of the disc. The discs around intermediate mass stars are warmer and remain warmer for longer, meaning key carbon and oxygen carriers remain the gas phase for a longer time period and in a greater extent of the disc. This dichotomy in disc composition has direct implications for the composition of planets that form in these areas. For example by 2~Myr in our models the CO snowline of discs around stars of 0.5 or 1~M$_\odot$ are within 40~au. We therefore do not expect giant planets rich in carbon to be found at wide separations in the discs around the lower mass stars. On the other hand, the discs around intermediate mass stars are much more likely to host carbon-rich giant planets on wider orbits because CO remains in the gas for a much greater separation in the disc; over 250~au in the case of the 3~M$_\odot$ star at >2~Myr. 

By applying our modelling to the HR8799 system we have been able to constrain a period of time in which the bulk of C- and O- carrying molecules must have been accreted to the two outer planets from the model disc. The position of the CO snowline rules out the formation of HR8799b beyond 1~Myr.
This result assumes that atmospheric C/O reflects that of the initial disc material, but chemical processes in exoplanetary atmosphere will alter the observed ratio. Furthermore modelling by \citep{Drozdovskaya2016CometaryMidplanes} shows that the initial composition of the disc is altered from that of the parent molecular cloud due to infall during the formation process, particularly in environments of temperature below 20~K where the balance of CO, CO$_2$ and H$_2$O ices changes to make CO$_2$ more abundant.  More precise predictions could be made by implementing a more sophisticated chemical network and using species beyond the main C and O carriers as tracers of atmospheric composition. 

In the case of HR8799 our results favour early accretion of C and O from the disc. These early stages can potentially include many other processes that can affect initial chemical composition that stem from disc formation mechanisms \citep{Drozdovskaya2016CometaryMidplanes} and dynamics in the disc,  especially if there is gravitational instability \citep{Ilee2011ChemistryDisc,Ilee2017,Evans2015GravitationalChemistry}.

In our analysis we consider in-situ formation of the giant planets in the system. An evolving temperature structure will also impact upon migration timescales of planets in the disc however, as migrational velocity of a planet increases proportional to temperature. In discs that cool with time as a result of stellar evolution, planets will migrate at a greater velocity at earlier times compared to the velocity of a planet of equal mass at the same radial separation at late times. In discs that reheat, CO is present in the gas phase for a greater proportion of the disc but any migrating planets within the disc will do so with a greater velocity than at early times in a cooler disc. If planet-driven gaps in the disc begin to overlap, and the planets themselves are in mean-motion resonance then convergent migration can occur \citep{Kley2000OnProtoplanets}, potentially helping to achieve mean motion resonances that stabilise the orbits of the planets \citep{Fabrycky2010STABILITYMASSES}.


\subsection{Implications for dust growth to pebble sizes}

In Section \ref{sec:HR8799} the analysis of HR8799c suggested that significant enrichment by solids could be required to explain the approximately solar C/O of the wide orbit planet. The amount of solids accreted depends on the method of the planet's formation. Here we explore the impact of stellar evolution on the composition of midplane dust, in particular we look to assess to what extent grain size in the disc is limited by the evolution of the star and whether this could prevent pebble accretion in regions of the disc. In order for pebble accretion to be rapid, pebbles have to make up a significant fraction for the gas giant core's solid density \citep{Lambrechts2012RapidAccretion}, and so they likely contribute significantly to oxygen enrichment of the forming planet. The term `pebbles' is used to refer to large grains that drift rapidly. In order to make quantitative comparisons in our models we will approximate this to mean dust grains with $a\geq$ 1~cm. 

\begin{figure}
    \centering
    \includegraphics[width=1\linewidth]{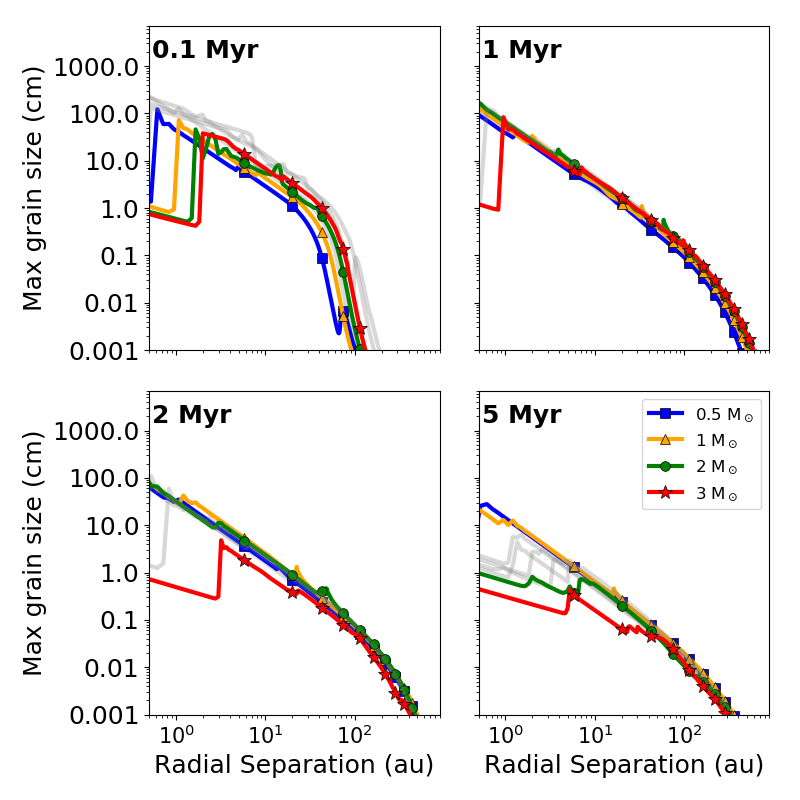}
    \caption{In each panel the maximum grain size is plotted as a function of radius for disc models at different time steps. At each time step the results are shown for disc models with different masses of host star plotted as different colours with markers as given in the key. An indication of the results for a model using stellar parameters that do not evolve, with values adopted from 1~Myr, are plotted in grey. }
    \label{fig:max_grains} 
\end{figure}

In Figure \ref{fig:max_grains} results of dust evolution in our models are plotted as coloured lines, the grey lines in the background show the results of modelling where the stellar parameters remain constant but the dust is still allowed to evolve. As expected an increase in grain size in the inner disc occurs due to changes in fragmentation velocity, e.g. the sharp jumps at <2~au for all the models at 0.1~Myr (Figure \ref{fig:max_grains}).   

Maximum grain size as a function of radius is similar among all models at 1~Myr, it is not until the divergence of midplane temperatures that a significant affect is seen in outer regions. The threshold between icy and ice-free fragmentation is pushed further into the disc as a result of the stellar evolution; this is most clearly seen by comparing the red/star lines showing the 3~M$_\odot$ models at different times in Figure \ref{fig:max_grains}. The extent to which stellar evolution is dictating this location can be seen by comparing to the similar radial features of the grey lines in the background.

In Figure \ref{fig:max_grains} grains are smaller than 1~cm in the innermost regions of the discs at 0.1~Myr, as the disc midplanes are relatively warm due to the luminous young star. This is a short lived stage however, as stellar luminosity evolves rapidly and the effects of radial drift start to take effect.  The models of discs around low mass stars at 1 and 2~Myr (blue/square and gold/triangle lines in Figure \ref{fig:max_grains}) host pebbles out to 20-30~au. By 5~Myr the 0.5 and 1 ~M$_\odot$ models still host pebbles as far as 10~au. These models show that pebbles reside in the inner disc ($<$10~au) for at least 5~Myr and will be able to contribute towards core growth in these regions. 
This is not the case in the discs around intermediate mass stars. The 1~Myr panel in Figure \ref{fig:max_grains} shows that fragmentation limit in the 3~M$_\odot$ model is found at 1~au, and is at 5~au in the 5~Myr panel. Due to rapid radial drift and fragmentation in collisions any pebbles in these inner regions will be removed from the disc quickly. Pebble sized grains are only efficiently removed from the inner regions of the 2~M$_\odot$ model by around 5~Myr when the effects of radial drift become dominant. This is signalled by the decrease of maximum grain size at all radial locations $\leq$10~au between 2 and 5~Myr, rather than the sharp jump associated with the fragmentation threshold.

The dust modelling presented in Figure \ref{fig:max_grains} thereby places some constraints on the ability of discs to host pebble sized grains as a result of the stellar luminosity evolution. Due to the cool discs of low mass stars, their discs remain virtually unaffected. After 1-2~Myr, fragmentation limits grains from growing to pebble sizes in intermediate mass discs as a result of stellar evolution and its subsequent impact on heating of the disc. This only applies within the first 1-5~au however and by $\approx$5~Myr radial drift becomes the dominant factor in removing large grains from the inner disc. Lack of large grains at short radial separation does not affect the formation of the HR8799 planets considered here, as they are on wider orbits. This would, however, affect the formation of giant planets forming at shorter separation, i.e. hot Jupiters, in planet forming discs.

\begin{figure}
    \centering
    \includegraphics[width=1\linewidth]{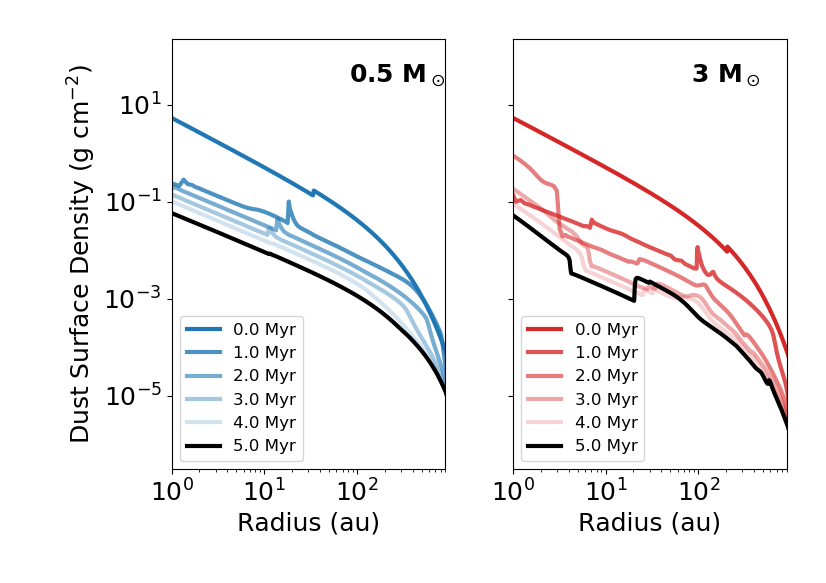}
    \caption{Surface density profile as a result of dust evolution in disc models with central stars of mass 0.5 and 3~M$_\odot$ for timesteps of 1~Myr. }
    \label{fig:sigma_graingrowth}
\end{figure}

An additional increase in grain size is seen in the models of discs around intermediate mass stars, where the CO$_2$ snowline creates a similar transition. For example at around 6~au in the 3~M$_\odot$ model at 5~Myr and at 30~au in the 3~M$_\odot$ model in the same panel of Figure \ref{fig:max_grains}. Unlike at the H$_2$O snowline there is not a strong effect on the fragmentation threshold \citep{Musiolik2016FORMATION}. Instead, the dust mass drops as the CO$_2$ leaves the grains, leading to a decrease in local dust surface density. In Figure \ref{fig:sigma_graingrowth} dust surface density is plotted for models at the extremes of our adopted stellar mass range, for regular time steps. Comparing the 3~M$_\odot$ model at 5~Myr (right hand panel, black line in Figure \ref{fig:sigma_graingrowth}) with the corresponding model in Figure \ref{fig:max_grains}, there is a decrease in dust surface occurring at the same radial separation as the increase in grain size. As discussed in \citet{Boneberg2018TheTRAPPIST-1}, the smaller grains are more closely coupled to the gas, decreasing the speed of inward radial drift. As a result dust surface density rises interior to the H$_2$O snowline. The discs around lower mass stars do not show the second grain size change. Similarly the 0.5~M$_\odot$ dust surface density profile in Figure \ref{fig:sigma_graingrowth} is not affected in the same way as the higher mass case. 

Similar perturbations in grain size can be seen in the outer disc towards the CO snowline, but due to the lower density at greater radial separation, the features do not persist and are smoothed out by transport processes. Additionally much of the CO crossing the CO snowline is converted in to CO$_2$ under the assumptions of our chemical model, and so the inward flux of carbon is instead later released at the CO$_2$ snowline. 
Three small spikes can be seen on the red line in the 1~Myr panel of Figure \ref{fig:max_grains} corresponding to early snowline-induced disturbances.
Transport processes and the evolving snowline location in discs mean that there may be some displacement between these radial features and the snowline itself for any given time.

We focus here on the effects of disc transport and stellar evolution, and so we do not include a prescription for dust trapping in our dust model, a mechanism which acts to prevent the rapid removal of pebbles from the disc on short timescales \citep{Pinilla2012}. Accurate predictions of the timescales for pebble removal in protoplanetary discs will require such considerations. Circular, thin rings of dust that are consistent with dust trapping are common among high resolution observations of bright protoplanetary discs \citep{Dullemond2018}. It therefore seems that dust trapping mechanisms may also be common; further imaging of large samples of discs that are capable of resolving dust rings are necessary to confirm this. Nevertheless, modelling without pressure bumps does a good job in reproducing the evolution of disc radius, mass and accretion rates \citep{Rosotti2019TheDiffer,Sellek2020AMasses}.

The temporal constraints of our modelling suggest that the pre-main sequence evolution of the host star could be used to identify when key snowline-dependant planet formation mechanisms occur. For example numerical modelling by \citet{Schoonenberg2017PlanetesimalOut} shows that planetesimal formation by the streaming instability is aided by the diffusion and condensation at the water snowline. Furthermore snowlines may play an important role in the dust growth leading to self-induced dust traps at the water or even CO snowlines \citep{Vericel2020Self-inducedDiscs}. For a given proto-planet there will be a window of opportunity for processes such as these to have taken place, the length of which depends upon the spectral type of the star.

\section{Conclusions}

In this paper models of pre-main sequence stellar evolution across a range of masses are combined with radiative transfer modelling of protoplanetary discs to study the impact of stellar evolution on midplane temperature. The grid of radiative transfer models quantify the change in temperature for continuous discs of constant mass when irradiated by stars at different stages of their pre-MS evolution. The consequences of these changes in midplane temperature are discussed with specific attention to the composition of the disc and what this means for the planets that form there.
A summary of the main conclusions is listed below. 

\begin{itemize}

    \item At ages $>$2Myr, stellar evolution of stars with mass $>$1.5M$_\odot$ diverges significantly from that of the lower mass stars. The increase in luminosity results in an increase in midplane temperature within the discs they host. Disc models around intermediate mass stars are not only warmer than discs around low- or solar- mass stars, but they also remain at their warmest for a greater duration of the stars' pre-main sequence evolution.

    \item Applying this modelling procedure to the HR8799 system, it is found that accretion of carbon and oxygen by exoplanet HR8799b must have occurred at early times in the disc's evolution whilst the planet was still inside the CO snowline. A specific time constraint for any individual disc will however require an accurate disc mass determination with a reliable estimate of mass loss rate from the disc. 

    \item Stellar evolution can limit growth of dust grains to pebble sizes in the first few au for discs around intermediate mass stars, but only after a 1-2 Myr. Radial drift is the most efficient mechanism at limiting growth to pebble sized in the inner disc. 
\end{itemize}{}

\subsection{Future Prospects}

We present in this paper a grid of models demonstrating the variation in midplane temperature, and subsequently in composition, of a protoplanetary disc as a result of pre-main sequence evolution of the host star. In this paper we apply our modelling to the system of HR8799, the only system for which C/O has been constrained for directly imaged giant planets on wide orbits. A greater number of wide-separation planets to compare with is sorely needed. Observations with James Webb Space Telescope and ARIEL will provide spectroscopy of an increased number of exoplanets from which C/O can be retrieved. The determination of C/O ratios in exoplanetary atmospheres at large separation from the star, ideally at 10s of au from the host, are crucial to confirming the link between disc and planet. In Figure \ref{fig:Rco} we make an initial comparison of the temperature profiles calculated for our models with known N$_{2}$H$^{+}$ rings that trace the CO snowline position. At the moment only a small number of these observations exist, and only in the cases where N$_2$H$^{+}$ rings are thin and well-resolved can the snowline position be constrained accurately \citep{Qi2019ProbingEmission}. Observations of N$_2$H$^{+}$ across the stellar mass range and for a range of ages would allow for us to test our predictions of snowline movement. 

Finally, we decouple radiative transfer calculations and disc evolution in our modelling in order to study discs massive enough to form giant planets at ages of 5-10~Myr. Unfortunately current evolution models predict disc lifetimes much shorter than this due to the efficiency of photoevaporation, or due to short growth times that result in short radial drift times. Such shorter timescales disagree with observed evidence of massive discs with ages $\geq$5~Myr (HD100546, HD163296, TW Hya). 
An exciting future prospect would be to produce a model that evolves both the star and the disc. Viscous evolution, disc dynamics and icy grain compositions have been shown to also influence snowline positions and subsequently C/N/O ratios in the disc \citep{Piso2015C/OACCRETION,Piso2016THEDISKS}. Disc evolution, particularly at later stages is influenced strongly by photoevaporation and so modelling will require a thorough knowledge of disc evolution across the stellar mass range. Photoevaporation models including EUV, FUV and X-ray photoevaporation currently only exist for T Tauri stars and not for their Herbig counterparts. The development of such models for intermediate mass stars (Kunitomo et al. in prep) will enable future studies to consider the effect of chemistry within the disc as a function of stellar luminosity \citep[e.g.][]{Walsh2015TheRegime} whilst also taking into account the evolution of disc structures around evolving pre-main-sequence stars.

\section*{Acknowledgements}

JM acknowledges funding through the University Research Scholarship from the University
of Leeds. The work of OP is supported by the Royal Society Dorothy
Hodgkin Fellowship. RAB acknowledges support from the STFC consolidated grant ST/S000623/1. J.D.I. acknowledges support from the Science and Technology Facilities Council of the United Kingdom (STFC) under ST/R000549/1 and ST/T000287/1.

\section*{Data Availability}
The data underlying this article will be shared on reasonable request to the corresponding author.


\bibliographystyle{mnras.bst}
\bibliography{paper3.bib}



\appendix

\section{Relative mixing of C and O}
\label{sec:App_mixing}
\begin{figure}
    \centering
    \includegraphics[width=0.9\linewidth]{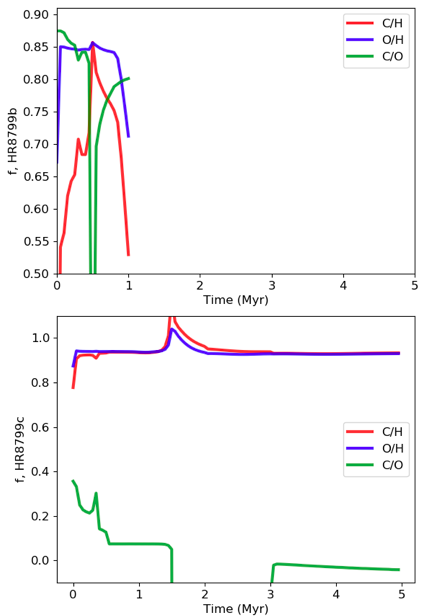}
    \caption{The proportion of material in the exoplanetary atmosphere that originated as gas phase disc material, calculated using Equation 2 in order to reproduce the retrieved atmospheric C/H, O/H and C/H for the exoplanets HR8799b (left) and HR8799c (right). The data in the left hand panel is plotted for all times at which the CO snowline is radially exterior to the HR8799b's orbital radius. }
    \label{fig:f_calc}
\end{figure}

Using Equation \ref{eqn:gas_frac} we can estimate how much of the exoplanetary atmosphere originated in the gas and solid phases through the relative amounts of carbon and oxygen in the atmosphere and in the disc as calculated by our modelling. Figure \ref{fig:f_calc} plots $f$ as a function of time through 5~Myr of the disc's evolution. CO dominates the gas phase material at the orbital radius of both planets, and so similar features can be seen in the C/H and O/H profiles. The top panel of Figure \ref{fig:f_calc} shows that most of the carbon and oxygen must originate from the gas phase at all simulation time. $f$ calculated through C/H approached a peak at around 0.5~Myr due to the inward drift of carbon-rich solids shown in Figure \ref{fig:HR8799}, varying between 0.5 and 0.85. O/H on the other hand remains fairly constant around 0.85, dropping at late times as the CO snowline approaches the planet. The combination of these two curves results in the green C/O line on the top of Figure \ref{fig:f_calc}.  At 0.5~Myr O/H is briefly greater than C/H in the disc model, due to an increase in the density of solids drifting in from the outer disc, resulting in an outlier results of very low f. Before and after this event $f$ calculated via C/O remains between $\approx$0.75-0.87. This calculation therefore does not support accretion from the inward drifting material that built up at a snowline in the outer disc, as the relative abundances cannot be reconciled with the observed C/O value. The build up of solids in the outer disc depends upon our assumptions of disc mass and $\alpha$. 

On the bottom of Figure \ref{fig:f_calc} $f$ is plotted for HR8799c as a function of time. $f$ calculated via C/H or O/H suggests that the vast majority of carbon and oxygen, $\geq 90\%$, must originate form the gas disc. This results in a high C/O, because gas is typically carbon rich and the solids oxygen rich. HR8799c has an approximately stellar detected C/O ratio however, leading to $f$ calculated via C/O that is less than 0.4 for all models. Between 1.5~Myr and 3~Myr no relative amounts of gas and dust can reproduce the measured C/O. This suggests that our formation assumptions do not apply here; namely that the composition of the exoplanetary atmosphere reflects the composition of the local gas and dust in the disc from which it formed. The discrepancy may be alerting us to significant chemical effects in the atmosphere, or to planet formation via gravitational instability.

\section{Alternative chemistry model}

In this paper our chemical model uses the temperature profile of the disc to establish the location of key snowlines. Volume mixing ratios and abundances of C and O are based upon observations of protoplanetary discs \citep{Draine2003,Pontoppidan2006SpatialCO2,Oberg2011} and carbon grains are included resulting in a greater amount of solid carbon. In previous studies of disc composition \citep[e.g.][]{Madhusudhan2014,Madhusudhan2017,Booth2017} this prescription has been compared to a similar model where volume mixing ratios are based on theoretical computations \citep{Woitke2009RadiationRim} and CH$_4$ is included in the species considered. As a result, the predicted disc C/O is altered, particularly in the dust and towards inner regions for the gas. 

\begin{figure} 
    \centering
    \includegraphics[width=1\linewidth]{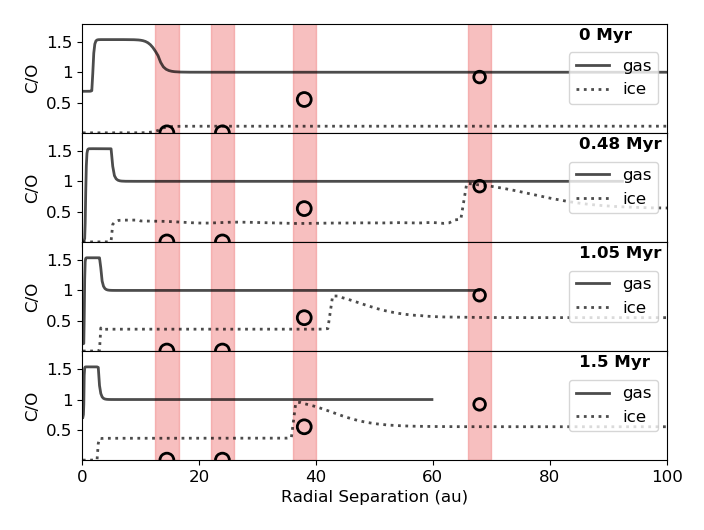}
    \caption{Alternative chemical model as presented in \citet{Madhusudhan2014}, and used as `Case 1' in \citet{Booth2017})}
    \label{fig:HR8799_Madhu}
\end{figure}

Table \ref{tab:madhu} gives the mixing ratios used for this model for comparison with Table \ref{tab:Oberg_chem}. Figure \ref{fig:HR8799_Madhu} shows results when this alternative chemistry model is applied to the disc model for HR8799. 

\begin{table} 
\begin{tabular}{cc}
\hline \hline
Species              & X/H \\ \hline
CO                     & 0.45 ( 1 + f$_{\rm CO_2}$) $\times$ C/H    \\
CH$_4$                 & 0.45 ( 1 - f$_{\rm CO_2}$) $\times$ C/H    \\
CO$_2$                 & 0.1 $\times$ C/H \\
H$_2$O                 & O/H - ( 3 $\times \frac{\rm Si}{\rm H}$ + $\frac{\rm CO}{\rm H}$ + 2 $\times$ $\frac{\rm CO_2}{\rm H}$ ) \\
Carbon grains        & 0  \\
Silicates            & Si/H  \\
\hline
\multicolumn{1}{l}{}    & \multicolumn{1}{r}{}                           
\end{tabular}
\caption{Binding energies, presented as temperatures, and volume mixing ratios of the key chemical species included in the alternative chemical model.}
\label{tab:madhu}
\end{table}

The key difference in the results from the alternative model comes within the CO$_2$ snowline, where due to the gas phase CH$_4$, C/O in the gas phase = 1.5, whereas solid phase material is down to 0, due to the lack of carbon grains. Although the radial profile of C/O in the disc is changed at short separation from the star, the alternative chemistry model has only a modest effect on the relative mixing ratios of gas and dust in order to reproduce the retrieved exoplanetary atmosphere values.


\bsp	
\label{lastpage}
\end{document}